\documentclass[english,twocolumn,aps,prl,showpacs]{revtex4}
\usepackage[T1]{fontenc}
\usepackage[latin1]{inputenc}
\usepackage{graphicx}
\usepackage{amssymb}

\makeatletter

\makeatletter

\makeatletter

\makeatletter

\input{epsf}

\makeatother

\makeatother

\makeatother

\usepackage{babel}
\makeatother
\begin{document}

\title{Phase diagram of a strongly interacting polarized Fermi gas in one
dimension}

\author{Hui Hu$^{1,2}$, Xia-Ji Liu$^{2}$, and Peter D. Drummond$^{2}$}

\affiliation{$^{1}$\ Department of Physics, Renmin University of China, Beijing
100872, China \\
 $^{2}$\ ARC Centre of Excellence for Quantum-Atom Optics, Department
of Physics, University of Queensland, Brisbane, Queensland 4072, Australia}

\date{\today{}}

\begin{abstract}
Based on the integrable Gaudin model and local density approximation,
we discuss the ground state of a one-dimensional trapped Fermi gas
with imbalanced spin population, for an arbitrary attractive interaction.
A phase separation state, with a polarized superfluid core immersed
in an unpolarized superfluid shell, emerges below a critical spin
polarization. Above it, coexistence of polarized superfluid matter
and a fully polarized normal gas is favored. These two exotic states
could be realized experimentally in highly elongated atomic traps,
and diagnosed by measuring the lowest density compressional mode.
We identify the polarized superfluid as having an FFLO structure,
and predict the resulting mode frequency as a function of the spin
polarization. 
\end{abstract}


\maketitle
Strongly attractive Fermi gases with imbalanced spin components are
common in different branches of physics~\cite{ketterle,hulet,muther,liu,sarma,bedaque,fflo,rmp},
so spin-polarized ultra-cold atomic gases are an atomic analog of
many other exotic forms of matter. The true ground state of an attractive
polarized Fermi gas remains elusive, because the standard Bardeen-Cooper-Schrieffer
(BCS) mechanism requires the pairing of two fermions with opposite
spin. Polarized Fermi gases cannot be explained by the BCS theory,
due to mismatched Fermi surfaces. Various exotic forms of pairing
have been proposed, including the deformed Fermi surface \cite{muther},
interior gap pairing \cite{liu} or Sarma superfluidity \cite{sarma},
phase separation \cite{bedaque}, and the inhomogeneous Fulde-Ferrell-Larkin-Ovchinnikov
(FFLO) state \cite{fflo}.

Recent measurements on polarized $^{6}$Li gases \cite{ketterle,hulet}
near a Feshbach resonance provide a route towards testing these theories
in experiment. However, the presence of a harmonic trap in these experiments
makes it difficult to identify which if any of the different pairing
schemes occurs. Theoretically, it is desirable to have an exactly
soluble mode of polarized uniform Fermi gases to identify various
pairing schemes, and clarify the issue of the trap.

In the present Letter we report on the exact ground state of a homogeneous
1D polarized Fermi gas~\cite{gaudin,takahashi,guan} with attractive
inter-component interactions at zero temperature. We then study the
phase diagram of an inhomogeneous Fermi gas under harmonic confinement,
using the local density approximation (LDA). We complement this with
a mean-field Bogoliubov-de Gennes (BdG) theory in the weak-coupling
limit, where the phase fluctuations are small, in order to clarify
the physical meaning of the solutions. Collective mode frequencies
are also calculated as an experimental diagnostic, thus extending
previous work on the unpolarized case~\cite{tokatly,zwerger,ldh}.

As well as being a theoretical test bed for the ground-state issue,
1D Fermi gases in traps can be realized using two-dimensional optical
lattices, where the radial motion of atoms is frozen out. Thus, one
can experimentally check these quantum many-body predictions, which
has also been recently carried out for the 1D Bose gas \cite{bosegas}.

The following remarkable features are found:

(\textbf{A}) In the ground state of a uniform system, we find three
distinct phases with increasing chemical potential difference between
species: an unpolarized BCS superfluid, a polarized superfluid, and
a fully polarized normal state. The polarized superfluid is most widespread,
and reduces to the FFLO-type for weak interactions. Therefore, it
is relatively easy to observe the FFLO physics in 1D, as anticipated
in previous approximate studies~\cite{1dpolarized}. In earlier work
the phase diagram was not conclusive, as the nature of the transition
from BCS to FFLO states was under debate~\cite{1dpolarized}.

(\textbf{B}) Within the local density approximation (LDA), we consider
the inhomogeneous phase diagram of the trapped gas. This leads to
a \emph{phase separation}, as  the inhomogeneous cloud separates into
either a mixture of a polarized superfluid and an unpolarized superfluid,
or a polarized superfluid and a fully polarized normal gas.

(\textbf{C}) We calculate the longitudinal size of the two spin components
and the frequency of the lowest density compressional mode. These
quantities give a measurable fingerprint of the whole phase diagram.

We describe a 1D polarized Fermi gas with $N=N_{\uparrow}+N_{\downarrow}$
fermions each of mass $m$ and spin polarization $P=(N_{\uparrow}-N_{\downarrow})/N>0$
in a harmonic trap, by \begin{equation}
{\cal H}={\cal H}_{0}+\sum\nolimits _{i=1}^{N}\frac{1}{2}m\omega_{ho}^{2}x_{i}^{2},\label{hami}\end{equation}
 where \begin{equation}
{\cal H}_{0}=-\frac{\hbar^{2}}{2m}\sum_{i=1}^{N}\frac{\partial^{2}}{\partial x_{i}^{2}}+g_{1D}\sum_{i=1}^{N_{\uparrow}}\sum_{j=1}^{N_{\downarrow}}\delta(x_{i}-x_{j})\label{gaudin}\end{equation}
 is the Hamiltonian of Gaudin model of a spin 1/2 Fermi gas attracting
via a short range potential $g_{1D}\delta(x)$ \cite{gaudin}. The
coupling constant $g_{1D}$ ($<0$) can be expressed through the 1D
scattering length $a_{1D}$, $g_{1D}=-2\hbar^{2}/(ma_{1D})$. A two-fermion
bound state arises once $N_{\downarrow}>0$, with binding energy $\epsilon_{b}=\hbar^{2}/(ma_{1D}^{2})$.

In the absence of the harmonic trap, the integrable Gaudin model,
Eq. (\ref{gaudin}), can be solved exactly using Bethe's ansatz \cite{gaudin,takahashi}.
Introducing linear number densities, $n=N/L$ and $n_{\sigma}=N_{\sigma}/L$
($\sigma=\uparrow,\downarrow$), where $L$ is the size of the system,
the uniform gas is characterized by the polarization $p=(n_{\uparrow}-n_{\downarrow})/n>0$
and a dimensionless parameter $\gamma=-mg_{1D}/(\hbar^{2}n)=2/(na_{1D})>0$.
The ground state is obtained by numerically solving the Gaudin integral
equations \cite{gaudin,takahashi}. We have clarified the physical
nature of the resulting solutions by also solving the weak-coupling
mean-field BdG equations.

%
\begin{figure}
\begin{centering}\includegraphics[clip,width=0.45\textwidth]{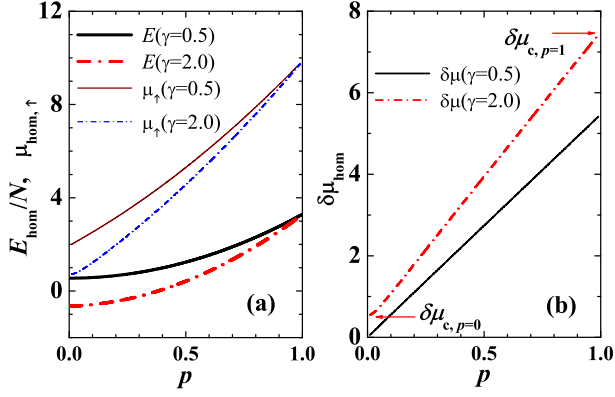}\par\end{centering}

\caption{(Color online) The energy per particle, chemical potential for spin
up atoms (a), and chemical potential difference (b) in units of $\hbar^{2}n^{2}/2m$
as a function of the spin polarization at two coupling parameters
$\gamma=0.5$ and $\gamma=2.0$.}

\label{fig1} 
\end{figure}


Fig. 1 shows the energy per particle, $E_{\hom}/N$, the chemical
potential of spin up fermions, $\mu_{\hom,\uparrow}=\partial E_{\hom}/\partial N_{\uparrow}$,
and the chemical potential difference, $\delta\mu_{\hom}=\partial E_{\hom}/\partial(N_{\uparrow}-N_{\downarrow})$,
as a function of polarization $p$ at two intermediate interaction
strengths $\gamma=0.5$ and $\gamma=2.0$. The mean-field calculations
lead to the same general behavior. Their asymptotic behavior in the
weak and strong coupling limits may be understood analytically. Of
particular interest is the chemical potential difference. In the weakly
interacting limit of $\gamma\ll1$, $\delta\mu_{\hom}$ is given by
($\gamma\ll\max\{ p,1-p\}$) \begin{equation}
\delta\mu_{\hom}=\frac{\hbar^{2}n^{2}}{2m}\frac{\pi^{2}}{2}p+\frac{\hbar^{2}n^{2}}{2m}\gamma p+\cdots,\label{dmuBCS}\end{equation}
 where the first term on the right hand side is the Fermi energy difference
of an ideal polarized gas, and the second term arises from the mean-field
Hartree-Fock interactions. The chemical potential difference increases
with increasing $\gamma$, and reaches a half of the binding energy
of bound states in the strongly attracting limit of $\gamma\rightarrow\infty$,
\begin{equation}
\delta\mu_{\hom}=\frac{\epsilon_{b}}{2}-\frac{\hbar^{2}n^{2}}{2m}\frac{\pi^{2}\left(1-p\right)^{2}}{16}+\frac{\hbar^{2}n^{2}}{2m}\pi^{2}p^{2}+\cdots.\label{dmuBEC}\end{equation}
 The first two terms in magnitude coincide with the chemical potential
of a Tonks-Girardeau bosonic gas of paired $N_{\downarrow}$ dimers
\cite{tokatly,zwerger}, which is fermionized due to strong attractions,
while the third term is equal to the chemical potential of residual
unpaired $N_{\uparrow}-N_{\downarrow}$ non-interacting fermions.
Therefore, in the strong coupling regime the system behaves like a
coherent mixture of a molecular Bose gas and fully polarized single-species
Fermi sea.

%
\begin{figure}
\begin{centering}\includegraphics[clip,width=0.4\textwidth]{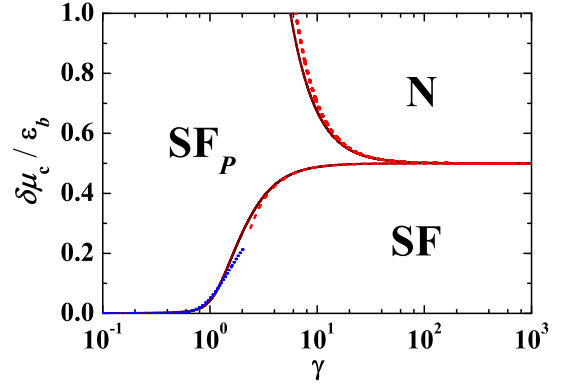}\par\end{centering}

\caption{(Color online) Uniform phase diagram, displaying {}``normal'' (N),
unpolarized superfluid (SF) and polarized superfluid states (SF$_{P}$).
At weak coupling or high density, the predominant SF$_{P}$ phase
corresponds to the mean-field FFLO solution . The dashed and dotted
lines are asymptotic behavior as described in the text.}

\label{fig2} 
\end{figure}


We analyze the phase structure of the ground state of uniform polarized
Fermi gases (Fig. 2). Given an interaction strength the chemical potential
difference at finite polarization $p$ is bounded by two critical
values, $\delta\mu_{c,p=0}$ and $\delta\mu_{c,p=1}$, as indicated
by arrows in Fig.1b for $\gamma=2.0$. Below $\delta\mu_{c,p=0}$,
the gas remains in the BCS-like superfluid state with zero polarization
(SF), while above $\delta\mu_{c,p=1}$, a fully polarized normal state
is favored (N). In between, a superfluid state with finite polarization
(SF$_{P}$) dominates. The mean-field calculation of a uniform gas
shows that the SF$_{P}$ is of FFLO character, as the exactly-soluble
ground state energy corresponds precisely with the FFLO solution in
the weak-coupling limit. Physically $\delta\mu_{c,p=0}$ is the energy
cost required to break pairs in unpolarized superfluid, \textit{i.e.},
the spin gap \cite{zwerger}, while $\delta\mu_{c,p=1}$ is also associated
with the pair-breaking (for the last pair), but is promoted upwards
due to the Pauli repulsion from existing fermions. The dependence
of $\delta\mu_{c,p=0}$ and $\delta\mu_{c,p=1}$ on the parameter
$\gamma$ is reported in Fig. 2, resulting in a homogeneous phase
diagram.

In units of $\hbar^{2}n^{2}/2m$, these have the following limiting
behavior: $\delta\mu_{c,p=0}\simeq 2\sqrt{\pi\gamma}\exp[-\pi^{2}/2\gamma]$
and $\delta\mu_{c,p=1}\simeq(\pi^{2}/2)$ as $\gamma\rightarrow0$,
while $\delta\mu_{c,p=0}\simeq\epsilon_{b}/2+\pi^{2}/16$ and $\delta\mu_{c,p=1}\simeq\epsilon_{b}/2+\pi^{2}$
as $\gamma\rightarrow\infty$. Both critical chemical potential differences
saturate to the half of the binding energy in the strong coupling
limit. Converting $\gamma$ into the chemical potential $\mu_{\hom}=\partial E_{\hom}/\partial N$,
we can obtain the phase diagram in the $\mu_{\hom}$-$\delta\mu_{\hom}$
plane.

%
\begin{figure}
\begin{centering}\includegraphics[clip,width=0.43\textwidth]{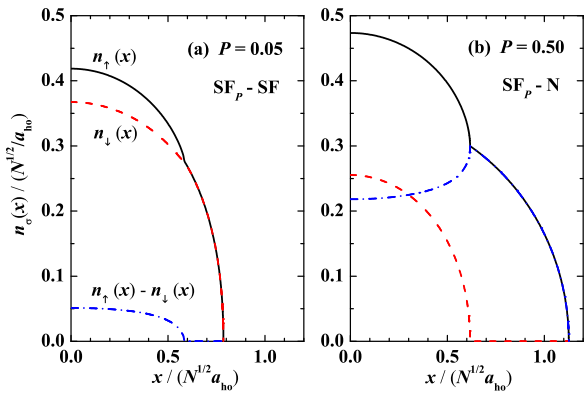}\par\end{centering}

\caption{(Color online) Density profiles of each spin component and their
difference at a typical interaction coupling of $Na_{1D}^{2}/a_{ho}^{2}=1$
for the SF$_{P}$-SF phase (a) and the SF$_{P}$-N phase (b).}

\label{fig3} 
\end{figure}


We now turn to describe the 1D polarized gas in a harmonic trap. We
partition the system into cells that can be treated locally as being
uniform, with a local chemical potential. This local density approximation
(LDA) is applicable provided that the number of fermions in a cell
is much greater than unitary, and the variation of the trap potential
across the cell is small compared with the local Fermi energy and
hence the interface effects are negligible~\cite{ldabreakdown}.
Overall it requires a sufficient large local density, which implying
$N\gg1$, a condition readily satisfied in the 1D experiment. Note
that the breakdown of LDA has been observed in the elongated 3D trap~\cite{ldabreakdown},
when the linear density in the transverse axis becomes small. The
LDA amounts to determining the chemical potential $\mu=(\mu_{\uparrow}+\mu_{\downarrow})/2$
and the chemical potential difference $\delta\mu=(\mu_{\uparrow}-\mu_{\downarrow})/2$
of the inhomogeneous gas from the local equilibrium condition, \begin{equation}
\mu_{\hom,\sigma}\left[n(x),p(x)\right]+\frac{1}{2}m\omega_{ho}^{2}x^{2}=\mu_{\sigma}.\label{LDA}\end{equation}
 The normalization conditions are $N=\int n(x)dx$ and $NP=\int n(x)p(x)dx$,
where $n(x)$ is the total linear number density and $p(x)$ the local
spin polarization. By rescaling the chemical potentials, coordinate
and linear density into dimensionless form {[}$\tilde{\mu}_{\sigma}=\mu_{\sigma}/\epsilon_{b}$,
$\tilde{x}=x/(a_{ho}^{2}/a_{1D})$ and $\tilde{n}=na_{1D}$], the
normalization equations can be rewritten as $Na_{1D}^{2}/a_{ho}^{2}=\int\tilde{n}(\tilde{x})d\tilde{x}$
and $(Na_{1D}^{2}/a_{ho}^{2})P=\int\tilde{n}(\tilde{x})p(\tilde{x})d\tilde{x}$.
These expressions emphasize that the dimensionless coupling constant
in a trap is controlled by $Na_{1D}^{2}/a_{ho}^{2}$, where $Na_{1D}^{2}/a_{ho}^{2}\gg1$
corresponds to weak coupling while $Na_{1D}^{2}/a_{ho}^{2}\ll1$ corresponds
to the strongly interacting regime.

The qualitative feature of density profiles $n_{\sigma}(x)$ is simple
to understand. Within the LDA, the local chemical potential $\mu(x)$
decreases parabolically away from the center of the trap while the
local chemical potential difference $\delta\mu(x)$ stays constant.
It is then clear from Fig. 2 that with a nonzero $P$ we always have
a polarized superfluid in the center where the local chemical potential
(interaction parameter) is large (small). Away from the center with
decreasing local chemical potential, the Fermi gas goes into either
an unpolarized superfluid ($\delta\mu<\epsilon_{b}/2$) or a fully
polarized normal cloud ($\delta\mu>\epsilon_{b}/2$). Thus, there
is a critical chemical potential difference $\delta\mu_{c}\equiv\epsilon_{b}/2$
that separates the inhomogeneous system into two phase separation
states: a mixture of a polarized superfluid core and an unpolarized
superfluid shell (SF$_{P}$-SF), or a coexistence of a polarized superfluid
at the center and a fully polarized normal gas outside (SF$_{P}$-N).

%
\begin{figure}
\begin{centering}\includegraphics[clip,width=0.45\textwidth]{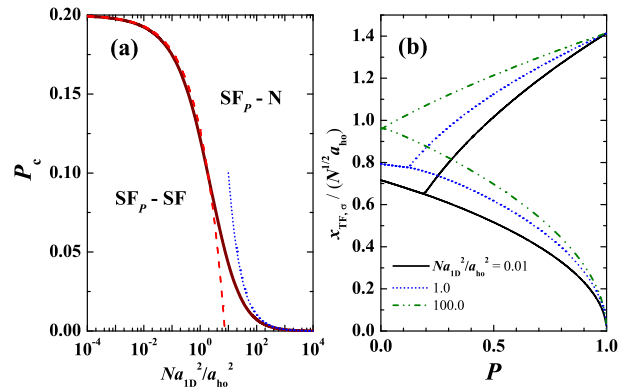}\par\end{centering}

\caption{(Color online) (a) Phase diagram of a inhomogeneous 1D polarized
Fermi gas. The dashed and dotted lines are asymptotic limits described
in the text. (b) The Thomas-Fermi radius of each spin component as
a function of the polarization at $Na_{1D}^{2}/a_{ho}^{2}=0.01$,
$1$, and $100$.}

\label{fig4} 
\end{figure}


The former phase is exotic, as the BCS-like superfluid state occurs
at the edge of the trap, in marked contrast to the 3D case \cite{ketterle}.
This is caused by the peculiar effect of low dimensionality, for which
the gas becomes more nonideal with \emph{decreasing} 1D density towards
the edge of the trap, and hence the energy required to break the pairs
approaches $\epsilon_{b}/2$ from below. As $\delta\mu<\epsilon_{b}/2$,
there should be a fully paired region once the local critical chemical
potential $\delta\mu_{c,p=0}>\delta\mu$, \textit{i.e.}, the BCS-like
superfluid.

We show the density profile of each component in Fig. 3 with a typical
coupling parameter $Na_{1D}^{2}/a_{ho}^{2}=1$. In addition, we have
performed a BdG calculation for the trapped gas and observe a FFLO-N
phase. The resulting density profiles on the weak coupling side are
in perfect agreement with the LDA calculation, indicating unambiguously
that the SF$_{P}$ phase is a FFLO state.

We determine the phase diagram of the inhomogeneous polarized 1D Fermi
gas as a function of the interaction strength and spin polarization
by calculating the critical polarization $P_{c}$ that corresponds
to $\delta\mu_{c}$ as a function of the coupling constant, and plot
it in Fig. 4a. The asymptotic behavior of $P_{c}$ can be computed
analytically in the weak and strong coupling limits. We find that
$P_{c}\simeq1/(Na_{1D}^{2}/a_{ho}^{2})$ as $Na_{1D}^{2}/a_{ho}^{2}\rightarrow\infty$,
and $P_{c}\simeq1/5-(256/225\pi^{2})(0.4Na_{1D}^{2}/a_{ho}^{2})^{1/2}$
as $Na_{1D}^{2}/a_{ho}^{2}\rightarrow0$.

We consider the experimental relevance of the two phase separation
states, by calculating the size of the cloud and the lowest density
compressional mode. These are readily detectable via absorption imaging.
Fig. 4b reports the evolution of the Thomas-Fermi radius of two spin
components as a function of polarization at three different interaction
couplings. The radius for spin up and down fermions is the same in
the SF$_{P}$-SF phase, but diverges in opposite directions in the
SF$_{P}$-N phase.

%
\begin{figure}
\begin{centering}\includegraphics[clip,width=0.4\textwidth]{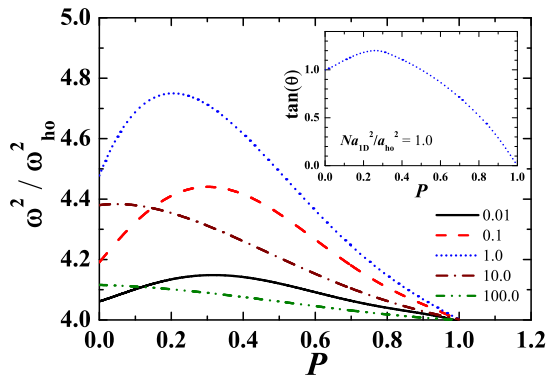}\par\end{centering}

\caption{(Color online) Square of the lowest density breathing mode frequency,
$\omega^{2}$, as a function of the spin polarization for interaction
parameters as indicated. The inset shows the mixing angle that minimizes
the sum-rule mode frequency at $Na_{1D}^{2}/a_{ho}^{2}=1.0$.}

\label{fig5} 
\end{figure}


Using a sum rule approach, the frequency $\omega$ of the lowest density
and spin density compressional (breathing) modes of 1D trapped gases
can be calculated from the identity $\hbar^{2}\omega^{2}=m_{1}/m_{-1}$,
where $m_{1}=\left\langle \left[F^{+},\left[{\cal H},F\right]\right]\right\rangle /2$
and $m_{-1}=-\chi(F)/2$ are two energy-weighted moments of the breathing
operator $F=\cos\theta\sum_{i}x_{i\uparrow}^{2}+\sin\theta\sum_{j}x_{j\downarrow}^{2}$,
with a mixing angle $-\pi/2<\theta<\pi/2$. The linear static response
of the system $\chi$ can be calculated in terms of the susceptibility
matrix $\partial^{2}E_{\hom}/\partial n_{\sigma}\partial n_{\sigma^{\prime}}$.
Basically these two different breathing modes correspond to the in-phase
and out-of-phase oscillations of the two imbalanced spin populations.
They are decoupled for a mixing angle $\theta_{in}>0$ or $\theta_{out}<0$
that minimizes the mode frequency, analogous to the spin-charge separation
in 1D. Then, the operators $F$ with these angles are anticipated
to exhaust all the weights in the density and spin density channels,
respectively. Thus, the sum rule approach is well applicable, providing
only an upper bound on the mode frequency. A stringent test of this
single mode approximation merits further study, \textit{e.g.}, by
using the random-phase approximation theory. Fig. 5 shows the density
breathing mode frequency as a function of polarization at several
interaction strengths. With increasing polarization the frequency
initially rises in the SF$_{P}$-SF state and then gradually decreases
to the ideal gas result $2\omega_{ho}$ in the SF$_{P}$-N phase.
A peak structure is found that gives an independent means of identifying
the FFLO phase, which dominates the phase diagram in the vicinity
of this peak.

We emphasize that the FFLO superfluid reported here can be detected
via a Josephson junction that is formed by confining two 1D polarized
gases in a double-well trap. There is also a signature present in
the density correlation function, which has an oscillatory behaviour
in this phase. Further details will be provided elsewhere.

In conclusion, we have investigated a 1D polarized Fermi gas in a
harmonic trap, and have shown that the trap generally gives rise to
phase separation: with at least one FFLO-type phase present. Two distinct
exotic phase-separated structures can occur, and are detectable via
absorption imaging and collective mode experiments.

We acknowledge fruitful discussions with Dr. X.-W. Guan. This work
is supported by the Australian Research Council Center of Excellence
and by the National Science Foundation of China under Grant No. NSFC-10574080
and the National Fundamental Research Program under Grant No. 2006CB921404.

\textit{Note added}.--- After this work was completed, we became aware
of a related work of G. Orso~\cite{orso}.


\begin{thebibliography}{10}
\bibitem{ketterle}M. W. Zwierlein \textit{et al.}, Science \textbf{311},
492 (2006); Nature (London) \textbf{442}, 54 (2006); Y. Shin \textit{et
al.}, Phys. Rev. Lett. \textbf{97}, 030401 (2006).

\bibitem{hulet}G. B. Partridge \textit{et al.}, Science \textbf{311},
503 (2006).

\bibitem{muther}H. Müther and A. Sedrakian, Phys. Rev. Lett. \textbf{88},
252503 (2002).

\bibitem{liu}W. V. Liu and F. Wilczek, Phys. Rev. Lett. \textbf{90},
047002 (2003).

\bibitem{sarma}G. Sarma, J. Phys. Chem. Solids \textbf{24}, 1029
(1963); C.-H. Pao, S.-T. Wu, and S.-K. Yip, Phys. Rev. B. \textbf{73},
132506 (2006); H. Hu and X.-J. Liu, Phys. Rev. A \textbf{73}, 051603(R)
(2006).

\bibitem{bedaque}P. F. Bedaque, H. Caldas, and G. Rupak, Phys. Rev.
Lett. \textbf{91}, 247002 (2003).

\bibitem{fflo}P. Fulde and R. A. Ferrell, Phys. Rev. \textbf{135},
A550 (1964); A. I. Larkin and Y. N. Ovchinnikov. Zh. Eksp. Teor. Fiz.
\textbf{47}, 1136 (1964) {[}Sov. Phys. JETP \textbf{20}, 762 (1965)].

\bibitem{rmp}For reviews, see for example, R. Casalbuoni and G. Nardulli,
Rev. Mod. Phys. \textbf{76}, 263 (2004).

\bibitem{gaudin}M. Gaudin, Phys. Lett. \textbf{24A}, 55 (1967).

\bibitem{takahashi}M. Takahashi, Prog. Theor. Phys. \textbf{44},
348 (1970).

\bibitem{guan}M. T. Batchelor \textit{et al.}, Journal of Physics
Conference Series \textbf{42}, 5 (2006).

\bibitem{tokatly}I. V. Tokatly, Phys. Rev. Lett. \textbf{93}, 090405
(2004).

\bibitem{zwerger}J. N. Fuchs, A. Recati, and W. Zwerger, Phys. Rev.
Lett. \textbf{93}, 090408 (2004).

\bibitem{ldh}X.-J. Liu, P. D. Drummond, and H. Hu, Phys. Rev. Lett.
\textbf{94}, 136406 (2005).

\bibitem{bosegas}H. Moritz \textit{et al.}, Phys. Rev. Lett. \textbf{91},
250402 (2003); K. V. Kheruntsyan \textit{et al.}, \textit{ibid}. \textbf{91},
040403 (2003); P. D. Drummond \textit{et al.}, \textit{ibid}. \textbf{92},
040405 (2004).

\bibitem{1dpolarized} K. Yang, Phys. Rev. B \textbf{63}, 140511 (R)
(2001); H. Shimahara, Phys. Rev. B \textbf{50}, 12760 (1994).

\bibitem{ldabreakdown} T. N. De Silva and E. J. Mueller, Phys. Rev.
Lett. \textbf{97}, 070402 (2006); A. Imambekov \textit{et al.}, Phys.
Rev. A \textbf{74}, 053626 (2006).

\bibitem{orso} G. Orso, Phys. Rev. Lett. \textbf{98}, 070402 (2007). 
\end{thebibliography}
\end{document}